# ALBANIAN LANGUAGE IDENTIFICATION IN TEXT DOCUMENTS

*KLESTI HOXHA.[1], ARTUR BAXHAKU.[2]

[1]University of Tirana, Faculty of Natural Sciences, Department of Informatics
[2]University of Tirana, Faculty of Natural Sciences, Department of Mathematics

email: klesti.hoxha@fshn.edu.al

**Abstract**

In this work we investigate the accuracy of standard and state-of-the-art language identification methods in identifying Albanian in written text documents. A dataset consisting of news articles written in Albanian has been constructed for this purpose. We noticed a considerable decrease of accuracy when using test documents that miss the Albanian alphabet letters "Ë" and "Ç" and created a custom training corpus that solved this problem by achieving an accuracy of more than 99%. Based on our experiments, the most performing language identification methods for Albanian use a naïve Bayes classifier and n-gram based classification features.

**Përmbledhje**

Në këtë punim shqyrtohet saktësia e disa metodave standarde dhe bashkëkohore në identifikimin e gjuhës shqipe në dokumente tekstuale. Për këtë qëllim është ndërtuar një bashkësi të dhënash testuese e cila përmban artikuj lajmesh të shkruara në shqip. Për tekstet shqipe që nuk përmbajnë gërmat "Ë" dhe "Ç" u vu re një zbritje e konsiderueshme e saktësisë së identifikimit të gjuhës. Për këtë arsye u krijua një korpus trajnues i posaçëm që e zgjidhi këtë problem duke arritur një saktësi prej më shumë se 99%. Bazuar në eksperimentet e kryera, metodat më të sakta për identifikimin e gjuhës shqipe përdorin një klasifikues "naive Bayes" dhe veçori klasifikuese të bazuara në n-grame.

**Fjalëkyçe**: Identifikimi i gjuhës, klasifikimi i teksteve, përpunimi i gjuhës natyrore, gjuha shqipe.

**Introduction**

Language identification is the task of automatically identifying the language that a text document has been written. With the ubiquitous nature of the internet nowadays, plenty of information is available and daily updated on the web in different languages. This huge amount of facts and data is processed by various information retrieval systems (search engines, knowledge bases, recommender systems, etc.). Language identification is therefore a crucial step in many natural language processing pipelines.

The first approaches of language identification made use of the fact that common short words have different frequencies in each language. They were followed immediately by n-gram (sequence of n characters in a text) based approaches. The state-of-the-art language identification toolkits of nowadays have achieved an accuracy greater than 99%, therefore many authors consider that the language identification task has been solved. However, because most of the language identification methods are supervised ones that depend on preliminary training by creating language models, the best configuration settings for each language need still to be investigated.

Albanian is an indo-european language (Mallory & Adams, 1997) spoken by about 8 million people in the world. It is a native language for people living in Albania, Kosovo and ethnic Albanians living in Albania's surrounding countries. The limited amount of research works about the natural language processing of Albanian has hindered the availability of information retrieval systems that deal with texts written in Albanian. A typical example of a system that strongly depends on the availability of natural language processing tools is a knowledge base of facts present in news articles (Hoxha *et al*., 2016).

In this work we evaluate the performance of the most common language identification approaches in identifying Albanian in written texts. Even though most of them report a high accuracy in detecting Albanian, the experiments were done using "low noise" datasets that do not fully reflect the text sources generally available on the web. For example the Albanian alphabet letters "Ë" and "Ç" are commonly misspelled as "E" and "C", because Albanian layout keyboards are not very popular.

In the rest of this paper after giving an overview of the most common language identification approaches, the testing data corpus creation and experimental results are described. Results are reported in terms of accuracy, the identification speed has not been evaluated. Based on the achieved results, we also propose a different approach in creating the Albanian language models used by supervised language identification algorithms.

**Language identification approaches**

In this section we detail the most common language identification approaches reported in the literature.

**Common short words**

In an early work, Grefenstette (1995) describes a language identification approach based on the probability for a short word (i.e. proposition) to appear in a specific language. He created language profiles that consisted of the most common short words for the language in question. Short words (five characters or less) are extracted from the test text and the probability for it to be in a particular language is calculated based on the available language profiles. He achieved very good results (99% accuracy) for texts longer than 15 words, but as expected this method did not perform well for shorter texts because they contain fewer common short words.

**N-Gram based methods**

N-grams are sequences of N characters extracted from a text. Being more tolerant to spelling and grammatical errors in comparison to words (Martins and Silva, 2005), they are more appropriate for language identification of texts commonly available on the web. In this section we describe the most common n-gram based language identification methods.

**N-Gram frequency statistics based methods**

Cavnar and Trenkle (1994) published the seminal paper of this category of methods. Many further works are adoptions of their work. Their method makes uses of a language profile that consists of a n-gram frequency hash table in reverse order (the most common n-grams are on top). A similar profile is constructed for the test text. For each n-gram of the test text profile, is calculated how far out of place is it in comparison with the order of it in a language profile. The actual distance measure is the sum of these "out of place" values (Figure 1). This is done for each candidate language, choosing the language that has the smallest distance measure with the test document. They achieved an average accuracy of 99.8% for documents with more than 300 characters and 98.6% for shorter ones when using language profiles with the top 300 most frequent n-grams.

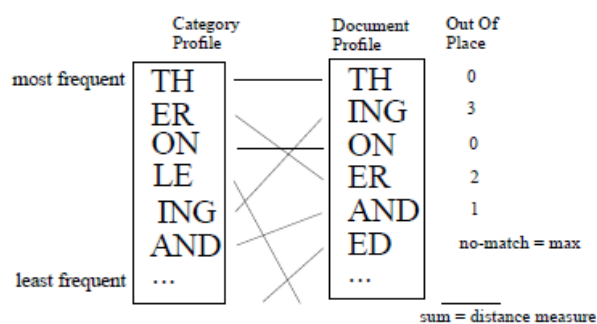

**Figure 1.** N-gram model comparison by Cavnar and Trenkle (1994)..

Martins & Silva (2005) use a modified version of the original Cavnar and Trenkle algorithm for identifying the language of web pages. Considering the nature of HTML structured documents they combine it with some heuristics that clean up the text, make use of meta data (if available) and weight n-grams based on the position of them in the document (title, descriptive meta tags, body). Numeric n-grams are also ignored. They performed experiments using the original rank order similarity measure and another similarity measure taken proposed by Lin (1998). Their results showed that the Lin similarity measure out-performed the original rank order statistics one in each experiment. The used heuristics also noticeably improved the achieved results. In average, their system achieved an accuracy of 99% in detecting Portuguese texts.

Ahmed *et al*. (2004) describe another modified version of the Cavnar and Trenkle algorithm. They use the Cumulative Frequency Addition (CFA) as a similarity measure instead of the rank order statistics one. Instead of ranking the test text n-gram profile frequencies, they just tokenize it (split in n-grams) and sum up the frequencies of each n-gram in the language profile in question.

If any n-gram does not exist in the language profile, it is simply ignored from the calculation. In this approach, n-grams may be found more than once in the test text profile (no frequencies are calculated). Therefore the computation of this profile is faster. Their results pointed out that the accuracy of the CFA similarity measure was comparable with the rank order statistics one. On the other hand, the computation speed for the CFA measure was 3-10 times faster.

**Machine learning N-Gram model methods**

One of the modern tools of language identification (Shuyo, 2014) uses a combination of a naïve Bayes classification approach with some normalization techniques for dealing with bias and noise in training and test corpus. It is offered as a JAVA library and its reported accuracy is 99.8% for 49 languages.

In (Lui & Baldwin, 2011) is described a state-of-the-art tool for language identification offered as a stand-alone Python module (Lui and Baldwin, 2012). It is a supervised machine learning approach that uses a multinomial naïve Bayes learner. Instead of representing documents as sequence of n-grams they use an information theoretic measure (Information Gain) over n-grams as the classification feature. They also try to reduce the amount of negative transfer learning, i.e. reducing the performance of the classifier when the training text data is from another domain. This is done by using training data from different domains and taking domains in consideration when selecting the classification features. They achieved an average accuracy of 99% and the classification speed is much faster in comparison to the other compared approaches. The experiments were conducted over a dataset of 97 languages from 5 different domains.

**Cosine similarity approach**

Brown (2013) describes a cosine similarity approach over a weighted subset of the most frequent n-grams of a language (based on the training data). He achieved very good results in identifying the language of short texts (at most 65 characters). The reported accuracy is 99.2% across 1100 languages when using a training model of 3500 n-grams.

**Bag-of-words based methods**

The bag-of-words model used in many information retrieval tasks, has also been applied to language identification. Zampieri (2013) reported on a machine learning approach that used bag-of-words as classification features. The achieved results were comparable to the traditional n-gram based approaches. He achieved an average accuracy of 96.8% when using a multinomial naïve Bayes classifier. He also pointed out that the bag-of-words method outperformed word unigram methods in identifying language variants (i.e. European and Brazilian Portuguese).

**Testing datasets**

In order to evaluate the performance of the language identification approaches in question, we have collected a testing data set of Albanian text documents that would indicate practical value in realistic conditions.

We collected news articles from ten news providers in Albania and one in Kosovo (publishes articles in the Gheg Albanian Dialect) covering multiple topics: politics, sports, showbiz, culture, economy, health, etc. This has been done by crawling their websites. For the news providers that published also news articles in English, we restricted the crawl only to the Albanian section of them (this was done based on the URL structure). The collected news articles were cleaned from duplicates and each of them was manually verified to be written in Albanian.

In order to allow the experimentation with text documents of various lengths in real settings, we extracted the title and content of each news article of the collection. In the following paragraphs of this section we describe in detail each dataset that was used for our experiments.

**Dataset 1 (D1)**: This dataset consists of 4575 articles of news providers in Albania. They are written in Standard Albanian (it is mostly based in Tosk dialect), with words being predominantly in the correct spelling and sentences obeying to the grammatical rules of the language.

**Dataset 2 (D2)**: This dataset simulates a common misspelling of the Albanian alphabet letters "Ë" and "Ç" due to the non general availability of Albanian layout keyboards. For this purpose we replaced all occurrences of those letters in Dataset D1 with "E" and "C" respectively.

**Dataset 3 (D3)**: In this dataset we wanted to simulate writers that occasionally use the letters "Ë" and "Ç". For this purpose we have probabilistically replaced the occurrences of these letters in Dataset D1 with "E" and "C" respectively. The replacement is done with a probability of 0.5.

**Dataset 4 (D4)**: This dataset consists of 500 byte excerpts from the contents of the articles of Dataset D1. Articles with contents length smaller than that were excluded. In total this dataset contains 4178 news articles.

**Dataset 5 (D5)**: This dataset consists of 2192 articles written in Gheg Albanian.

Table 1 gives an overview of the contents and title lengths of the news articles of the above described datasets.

**Table 1.** Length of News Article Title and Contents in Bytes

| DATASET | MIN | MAX | AVG |
|---|---|---|---|
| D1-D3 (Title) | 5 | 183 | 65 |
| D1-D3 (Contents) | 150 | 43988 | 2716 |
| D4 | 500 | 500 | 500 |
| D5 (Title) | 17 | 148 | 63 |
| D5 (Contents) | 150 | 25496 | 800 |

**Experimental setup and results**

Our experiments were conducted with open source toolkits that implement some of the approaches described in the previous sections. *TexCat*[1] is an open source implementation of the original Cavnar and Trenkle (1994) algorithm. *WhatLang*[2] is an open source implementation of Brown's cosine similarity based approach (Brown, 2014). *LangDetect*[3] is a Java library that uses a naïve Bayes n-gram based classification approach. l*angid.py*[4] (Lui & Baldwin, 2012) is a python stand-alone library that implements the language identification method described in (Lui & Baldwin, 2011).

The main aim is to find out which method and under which configuration achieves better accuracy in a dataset that resembles Albanian text documents found in the web. This would allow for a focused crawling of the "Albanian Web".

**Identify correctly spelled standard Albanian**

In this experiment we used the D1 dataset. The title and contents of each news article of this dataset was classified with the above mentioned open sources tools using their pre-trained language profiles (they contain a language profile for the Albanian language). The results of the experiments are found in Table 2. *TexCat* failed on identifying Albanian in most of the news article titles and also performed slightly worse than the other approaches in correctly identifying the contents language. *LangDetect* and *langid.py* achieved the same accuracy in detect-ing the language of the contents of the articles (99.96%) while *LangDetect* performed slightly better in identifying the language of the title.

---

[1] http://odur.let.rug.nl/~vannoord/TextCat/
[2] https://sourceforge.net/projects/la-strings/
[3] https://github.com/shuyo/language-detection
[4] https://github.com/saffsd/langid.py

**Albanian texts that miss the letters "Ë" and "Ç"**

In this experiment we used the D2 dataset. We used again the above mentioned open source tools for identifying the language of the title and contents of each news article. The results are displayed in Table 3.

Table 2. Accuracy in identifying correctly spelled standard Albanian

| TOOL | TITLE | CONTENTS |
| --- | --- | --- |
| LangDetect | 0.9593 | 0.9996 |
| langid.py | 0.9454 | 0.9996 |
| TexCat | 0.1657 | 0.9604 |
| WhatLang | 0.8997 | 0.9993 |

Table 3. Accuracy in identifying Albanian texts that miss the letters Ë and Ç

| TOOL | TITLE | CONTENTS |
| --- | --- | --- |
| LangDetect | 0.8323 | 0.9996 |
| langid.py | 0.6490 | 0.9987 |
| TexCat | 0.1170 | 0.9545 |
| WhatLang | 0.7233 | 0.9991 |

Results were clearly affected by the misspelling of the letters "Ë" and "Ç", especially the accuracy of identifying the title languages reduced by 20-30%. This shows that the accuracy of detecting Albanian language in short texts depends on the availability of these two letters in these texts (at least with the pre-trained language profiles).

**Albanian texts that miss some letters "Ë" and "Ç"**

In this experiment we used the D3 dataset. The titles and content of the news of this dataset contain misspelled letters "Ë" and "Ç" with a probability of 0.5. The results are displayed in Table 4. They verify the assumption of the experiment presented in the previous sections. The presence or missing of letters "Ë" and "Ç" clearly affects the accuracy of the language detection methods in question, especially when dealing with short texts.

**Fixed length test subjects**

In this experiment we used the D4 dataset. We examined the effects of the length of the test text in the accuracy of the language identification methods in question. Table 5 contains the results of this experiment. Interestingly enough the accuracy of three of the tools was slightly improved. Only the Cavnar and Trenkle based method (TexCat) accuracy dropped by 10%.

**Identifying gheg Albanian**

The Standard Albanian language is mostly based in its Tosk Dialect. Many words in Gheg (the other Albanian dialect) have different word endings or show different phonetic patterns. Gheg is also the dialect that Albanians in Kosovo speak. In this experiment we used the D5 dataset (contains news articles written in Gheg) for testing the accuracy of the tools in question in correctly identifying Gheg Albanian. Results are displayed in Table 6.

**Table 4.** Accuracy in identifying Albanian texts that miss some letters Ë and Ç

| TOOL | TITLE | CONTENTS |
| --- | --- | --- |
| LangDetect | 0.9339 | 0.9996 |
| Langid.py | 0.8879 | 0.9993 |
| TexCat | 0.1668 | 0.9515 |
| WhatLang | 0.8223 | 0.9991 |

**Table 5.** Accuracy on identifying Albanian texts 500 bytes long

| TOOL | TITLE | CONTENTS |
| --- | --- | --- |
| LangDetect | N/A | 0.9998 |
| Langid.py | N/A | 1.0000 |
| TexCat | N/A | 0.8621 |
| WhatLang | N/A | 0.9998 |

Only TexCat performed poorly in identifying Gheg Albanian. For the other three tools we noticed almost the same accuracy in comparison with the experiment on identifying Standard Albanian. The results of this experiment may be related to the training texts that were used for creating the in box language profiles of these tools (they might have contained some Gheg Albanian texts).

**Table 6.** Accuracy in identifying gheg Albanian

| TOOL | TITLE | CONTENTS |
| --- | --- | --- |
| LangDetect | 0.9772 | 0.9995 |
| Langid.py | 0.9567 | 0.9991 |
| TexCat | 0.1241 | 0.7509 |
| WhatLang | 0.9015 | 0.9986 |

**Custom trained langid.py experiments**

Based on the results of the above described experiments, we tried to improve the achieved accuracy by custom training the langid.py tool (together with LangDetect it was the most performing tool for most experiments). For this purpose we used the contents of 2000 articles of D3 as a training dataset.

The training corpus of langid.py needs to be organized in folders that contain text documents of different domains (Lui & Baldwin, 2012) for each involved language. We created a training dataset for Albanian and English (using also some English news articles) that contained news articles from these domains (topics): politics, economics, showbiz, food, sports, and technology. The total size of this corpus is 2000KB.

As testing datasets we used the articles of D2 that were not used for generating the training dataset from D3 (we call this dataset D2'). We also performed experiments using D5 (in order to test the performance of this language profile in identifying Gheg Albanian.

The results are displayed in Table 7. The custom training of the langid.py tool by using training text with some misspelled letters "Ë" and "Ç" greatly increased the accuracy of the Albanian language identification. The achieved accuracy was more than 99% for both Standard and Gheg Albanian.

**Table 7.** Achieved accuracy using the custom trained langid.py

| DATASET | TITLE | CONTENTS |
|---|---|---|
| D2' | 0.9981 | 1.0000 |
| D5 (Gheg Albanian) | 0.9991 | 1.0000 |

**Conclusion and Future Work**

Even though the language identification problem is considered as solved by some researchers, most of the available language identification methods have been evaluated by using non gold standard datasets that do not always reflect the nature of text documents encountered in the web in real scenarios.

In this paper we investigated the accuracy of four standard and state-of-the-art methods of language identification in detecting Albanian written text documents. Experiments were conducted using open source implementations of the methods in question.

We used a dataset consisting of news articles published in online news media in Albania and Kosovo written in Standard or Gheg Albanian. They cover different topics (domains) containing so a large lexicon that allows for simulation of real language identification scenarios. It needs to be mentioned that usually news articles go through an editing process, or are written by professional content writers. Therefore they most probably contain words in the correct spelling and sentences obeying to the grammatical rules of the language. For this reason we simulated a common misspelling of Albanian words in two of the used dataset variants.

Our results showed that the original Cavnar & Trenkle (2014) language identification algorithm fails on identifying the Albanian language in short texts (the achieved accuracy by this method was only 16%). The most performing tools in our experiments were LangDetect and langid.py. They achieved an accuracy of about 95% for short texts (news article titles in our case) and more than 99% for long texts. This was true for both Standard and Gheg Albanian written news articles. Both of these tools use a naïve Bayes classifier and n-gram based classification features.

In our experiments, we showed that the misspelling of the Albanian alphabet letters "Ë" and "Ç" as "E" and "C" highly affects the accuracy of the investigated tools to identify Albanian. This is especially true for shorter texts (accuracy dropped by 20-30%). In order to deal with this we experimented with a custom build training corpus that included random (with a probability of 0.5) misspelled versions of these letters. We trained langid.py with this corpus and achieved an accuracy of more than 99% for both Standard and Gheg Albanian.

In the future we plan to experiment with more noisy data like social media statuses that contain more misspelled words and grammatically incorrect sentences.

**References**

Ahmed, B., Cha, S., Tappert, C. (2004): Language Identification from Text Using N-gram Based Cumulative Frequency Addition. In: Proceedings of Student/Faculty Research Day, CSIS, Pace University: 1-8

Brown, R. D. (2013): Selecting and weighting n-grams to identify 1100 languages. In: Proceedings of International Conference on Text, Speech and Dialogue: 475-483

Cavnar W. B., Trenkle J. M. (1994): N-Gram-Based Text Categorization. In: Proc. SDAIR-94, 3rd Annu. Symp. Doc. Anal. Inf. Retr.: 161–175

Grefenstette G. (1995): Comparing Two Language Identification Schemes. In: Proc. 3rd Int. Conf. Stat. Anal. Textual Data (JADT 1995): 263–268

Hoxha K., Baxhaku A., Ninka I. (2016): Bootstrapping an Online News Knowledge Base. In: Proceedings of the 16th International Conference on Web Engineering (ICWE), Lugano, Switzerland: 501-506

Lin, D. (1998): An information-theoretic definition of similarity. In ICML, Vol. 98, No. 1998: 296-304

Lui, M., Baldwin, T. (2011): Cross-domain feature selection for language identification. In: Proceedings of 5th International Joint Conference on Natural Language Processing: 553-561

Lui M., Baldwin T. (2012): langid.py: An Off-the-shelf Language Identification Tool. In: Proceedings of the ACL 2012 System Demonstrations: 25–30

Mallory J., Adams D. Q. (1997): Encyclopedia of Indo-European Culture. Fitzroy Deadborn Publishers: 8-12

Martins, B., Silva, M. J. (2005): Language identification in web pages. In: Proceedings of the 2005 ACM symposium on Applied computing: 764-768

Shuyo N. (2014): Language Detection Library for JAVA. Available: https://github.com/shuyo/language-detection

Zampieri, M. (2013): Using bag-of-words to distinguish similar languages: How efficient are they?. In: Computational Intelligence and Informatics (CINTI), 2013 IEEE 14th International Symposium: 37-41